\documentclass[prb,twocolumn,amsmath,aps,byrevtex,groupaddress,subeqn]{revtex4}
\usepackage{graphicx}
\usepackage{epsfig}
\usepackage[cp1251]{inputenc}
\usepackage[english]{babel}
\usepackage{amsmath}
\usepackage{amssymb}
\usepackage{amstext}
\usepackage{color}
\usepackage[toc,page]{appendix}
\usepackage{comment}
\usepackage{epstopdf}

\begin{document}

\title{The temporal picture for Bloch electron dynamics in homogeneous electric fields}

\author{G. J. Iafrate}
\email[Corresponding author E-mail:\,]{gjiafrat@ncsu.edu}
\affiliation{Department of Electrical and Computer Engineering,
North Carolina State University, Raleigh, North Carolina 27695-8617, USA}
\author{V. N. Sokolov}
\affiliation{Department of Theoretical Physics,
Institute of Semiconductor Physics, NASU, Pr. Nauki 41, Kiev 03028,
Ukraine.}

\begin{abstract}
The transient picture for a Bloch electron accelerating in an arbitrarily time-dependent homogeneous electric field  is developed. The temporal sequence for the analysis includes the instant after electron injection, followed by the time required for a small change in electron wavenumber away from initial injection, leading to the final time evolution over many Bloch periods. The time-dependent behavior is studied using the properties of the Schr\"{o}dinger equation. The electric field is described through the vector potential gauge, and the instantaneous eigenstates of the Bloch, electric-field-dependent Hamiltonian are used as basis states in describing the Bloch dynamics in the electric field.
For each temporal sequence considered, the solution to the Schr\"{o}dinger equation is established and comparatively discussed. The expectation value of the momentum is obtained for the special case of first order in a constant electric field; the resulting velocity derived is a field-dependent generalization of the natural Zitterbewegung-like behavior discussed in the recent literature. The early-time and long-time limits of the momentum expectation value and its time derivative demonstrate that the resistance to Bloch acceleration after initial band injection varies from real mass to effective mass dynamics as the electron accelerates through the band under the influence of electric field. This changing inertia from early injection of a free-mass electron is the result of the {\it real mass} electron {\it dressing-up} into the states of the crystal to become {\it an effective mass} electron.
The ramifications of this temporal {\it dressing} behavior are discussed in considering the general dynamics of Bloch electrons subject to ultrastrong electric fields.
\end{abstract}

\maketitle

\section{Introduction}
Bloch electron dynamics has been a subject of central interest from the early development of solid state physics. The {\it early-time} behavior of Bloch electrons in electric fields is especially noteworthy and dates back to the original discussions concerning Bloch electron resistance to acceleration in electric fields~[\onlinecite{Krets,Peierls,Spenke}]. Today, with the advent of attosecond time-resolved experimental methods~[\onlinecite{Krausz,Ghimire}], the basic unfolding of the physical properties of Bloch electrons in intense time varying electric fields over a wide temporal range has emerged as an area of central interest. Therefore, in this effort, we systematically study the transient and temporal behavior of Bloch electron dynamics from initial electron injection into a crystal band through many Bloch periods where the electron is accelerating under an arbitrarily time-dependent, homogeneous electric field.

In Section II, we describe as introductory background the Bloch Hamiltonian for a homogeneous, arbitrarily time-dependent electric field, with the electric field described through the vector potential gauge (this choice of gauge preserves the crystal periodicity of the Hamiltonian and also provides a handle on the relationship between the electric field and its time evolution). From the time-dependent Schr\"{o}dinger equation, the complete set of instantaneous eigenstates (accelerated Bloch states) is developed, and then utilized as basis states to explore the temporal analysis.

In Section III, the Schr\"{o}dinger equation is utilized to establish the Bloch wave function for critical time spans relevant to the analysis. Three time spans are sequentially considered ranging from just after injection, then a $\delta k$ within the first cycle of the Brillouin zone, and, finally, many cycles of the Brillouin zone. The Bloch wave function for each time span is established and comparatively discussed. For each time span, the expectation value of the momentum (and its derivative) are calculated to first order in a constant electric field. The early-time and long-time limits of the momentum (time derivative) expectation values demonstrate that the resistance to Bloch electron acceleration after initial injection varies from real mass to effective mass dynamics as the electron propagates in the band under the influence of the electric field. This changing inertia from early injection of a free-mass electron is the result of the {\it real mass} electron {\it dressing-up} into the states of the crystal to become {\it an effective mass} electron. It is estimated that the transition from real mass to effective mass is the time to travel several lattice spaces, or {\it dress-up} into the states of the crystal. We note here that temporal effective mass dynamics has been discussed in connection with low-dimensional and optical lattices~[\onlinecite{Sipe1,Sipe2,Sipe3}], as well as in bulk GaAs~[\onlinecite{Zhu1}]. We also note that our derived expression for the oscillatory Bloch velocity is a field-dependent generalization of the natural {\it Zitterbewegung}-like behavior discussed by Zawadzki and Rusin~[\onlinecite{Zawadzki}] for semiconductors, and by Winkler et al.~[\onlinecite{Winkler}] for various Hamiltonians with gapped energy spectrum.

In Section IV, a summary of key results and conclusions are presented. Appendix A presents a matrix representation of the time evolution operator equation for the Bloch Hamiltonian of interest.

\section{Background: dynamics in a homogeneous electric field; Instantaneous eigenstates}
The Hamiltonian describing the dynamics~[\onlinecite{Krieger,Iafrate2}] for a Bloch electron in a crystal potential, $V_c({\bf r})$, subject to a time-dependent homogeneous electric field is
\begin{equation}\label{H}
 \hat{H}({\bf r},\hat{\bf p},t) = \frac{1}{2m_0}\big[\hat{\bf p} - \frac{e}{c} {\bf A}(t)\big]^2 + V_c({\bf r});
\end{equation}
here, $m_0$ is the free electron mass, $\hat{\bf p} = -i\hbar {\bf \nabla}_{\bf r}$ is the electron momentum operator,  $e$ is the electron charge ($e_0 = - e > 0$), $c$ is the speed of light in a vacuum, ${\bf r}$ is the space coordinate, and ${\bf A}({\bf r},t)$ is the vector potential given by
\begin{equation}\label{At}
{\bf A}(t) = -c\int_{t_0}^{t} {\bf E}(t^{\prime})dt^{\prime},
\end{equation}
where $t_0$ is the time that the electric field, ${\bf E}(t)$, is turned on. For convenience, we define
\begin{subequations}
\begin{equation}\label{pc}
 -\frac{e}{c}{\bf A}(t)  = e\int_{t_0}^t {\bf E}(t^\prime)dt^\prime \equiv {\bf p}_c(t),
\end{equation}
thus noting that
\begin{equation}\label{Ft}
 \dot{{\bf p}}_c  = {\bf F}(t),
\end{equation}
\end{subequations}
where ${\bf F}(t) = e{\bf E}(t)$ is the external force on the electron. We then insert Eq.~(\ref{pc}) into the Hamiltonian of Eq.~(\ref{H}) and write the
Schr\"{o}dinger equation for the dynamics as
\begin{equation}\label{sHr}
 i\hbar\frac{\partial}{\partial t} |\Psi({\bf r},t)\rangle = \Big[\frac{1}{2m_0}\big(\hat{\bf p} + {\bf p}_c(t)\big)^2 + V_c({\bf r})\Big] |\Psi({\bf r},t)\rangle.
\end{equation}

In seeking the solution to Eq.~(\ref{H}), we look to utilize the instantaneous eigenstates of the Schr\"{o}dinger equation, Eq.~(\ref{sHr}). This means we look to solve
\begin{equation}\label{H1}
 \Big[\frac{1}{2m_0}\big(\hat{\bf p} + {\bf p}_c(t)\big)^2 + V_c({\bf r})\Big] |\tilde{\Psi}({\bf r},t)\rangle = \varepsilon(t)|\tilde{\Psi}({\bf r},t)\rangle,
\end{equation}
where both $|\tilde{\Psi}({\bf r},t)\rangle$ and $\varepsilon(t)$ have an admitted time dependence due to presence of ${\bf p}_c(t)$.
In seeking the solution to Eq.~(\ref{H1}), we let
\begin{equation}\label{psiphi}
 |\tilde{\Psi}({\bf r},t)\rangle = e^{-i\frac{{\bf p}_c}{\hbar}\cdot{\bf r}} |\Phi({\bf r},t)\rangle;
\end{equation}
placing $|\tilde{\Psi}({\bf r},t)\rangle$ of Eq.~(\ref{psiphi}) into Eq.~(\ref{H1}), we find
\begin{equation}\label{phi}
 \Big[\frac{1}{2m_0}\hat{\bf p}^2 + V_c({\bf r})\Big] |\Phi({\bf r},t)\rangle = \varepsilon(t)|\Phi({\bf r},t)\rangle.
\end{equation}

We note that the solutions to Eq.~(\ref{phi}) are the Bloch functions and the energy-band funstions, but as a function of time through the time dependence of the wave vector, ${\bf k}(t)$. Therefore, we express $|\Phi({\bf r},t)\rangle$ and $\varepsilon(t)$ as
\begin{subequations}
\begin{equation}\label{phin}
 |\Phi({\bf r},t)\rangle = e^{i{\bf k}(t)\cdot{\bf r}} u_{n{\bf k}(t)}({\bf r})
\end{equation}
and
\begin{equation}\label{epsn}
 \varepsilon(t) = \varepsilon_n({\bf k}(t)),
\end{equation}
\end{subequations}
where $n$ is the energy-band index. To establish the allowed values of ${\bf k}(t)$, we note that the Hamiltonian of Eq.~(\ref{H1}) is invariant under lattice translation. Thus, imposing periodic boundary conditions on $|\tilde{\Psi}({\bf r},t)\rangle$ in Eq.~(\ref{psiphi}), we find
\begin{equation}\label{BC}
 {\bf k}(t) - \frac{1}{\hbar}{\bf p}_c(t) = \sum_{i=1}^3\frac{g_i}{N_i} {\bf G}_i,
\end{equation}
where ${\bf G}_i$ are the primitive reciprocal-lattice vectors ($i$ denotes a dimension), $N_i$ are the number of cells in the $i$ direction, and $g_i$ are integers with $-\frac{N_i}{2} < g_i < \frac{N_i}{2}$ .
It follows from Eq.~(\ref{BC}) that
\begin{subequations}
\begin{equation}\label{AcTh}
 \hbar \dot{\bf k}(t) = \dot{\bf p}_c(t) = {\bf F}(t),
\end{equation}
which establishes the {\it acceleration theorem}. Therefore, we can adopt the notation
\begin{equation}\label{Not1}
 {\bf k}(t) = {\bf K} + {\bf k}_c(t),
\end{equation}
where
\begin{equation}\label{Not2}
 {\bf k}_c(t) = \frac{1}{\hbar}\int_{t_0}^t {\bf F}(t^\prime)dt^\prime
\end{equation}
\end{subequations}
and ${\bf k}(t_0) = {\bf K}$.

Thus, the instantaneous eigenstates of Eq.~(\ref{psiphi}) are
\begin{equation}\label{psin1}
 |\tilde{\Psi}_{n{\bf K}}({\bf r},t)\rangle = \frac{1}{\sqrt{\Omega}}e^{i{\bf K}\cdot{\bf r}} u_{n{\bf k}(t)}({\bf r}),
\end{equation}
with $(\tilde{\Psi}_{n{\bf K}},\tilde{\Psi}_{n^\prime{\bf K}^\prime}) = \delta_{n,n^\prime} \delta_{{\bf K},{\bf K}^\prime}$,
whereas the states of Eq.~(\ref{phin}) are
\begin{equation}\label{phin1}
 |\Phi_{n{\bf k}}({\bf r},t)\rangle = \frac{1}{\sqrt{\Omega}} e^{i{\bf k}(t)\cdot{\bf r}} u_{n{\bf k}(t)}({\bf r}),
\end{equation}
where $\Omega$ is the crystal volume. These states are the so-called {\it Houston states}. In comparing $\Phi_{n{\bf k}}$ and $\tilde{\Psi}_{n{\bf K}}$, we note that $\Phi_{n{\bf k}}$ are a total function of ${\bf k}(t)$; whereas in $\tilde{\Psi}_{n{\bf K}}$, the instantaneous eigenstates, the exponential phase factor in $\tilde{\Psi}_{n{\bf K}}$ of Eq.~(\ref{psin1}) depends on ${\bf K}$ alone, independent of time, whereas the cellular Bloch function, $u_{n{\bf k}(t)}({\bf r})$, is dependent upon ${\bf k}(t)$, and therefore the electric field through Eq.~(\ref{Not1}).

In seeking the solution to the Schr\"{o}dinger equation of Eq.~(\ref{sHr}), we employ the complete set of instantaneous eigenstates of Eq.~(\ref{psin1}), $|\tilde{\Psi}_{n{\bf K}}\rangle$, to obtain
\begin{equation}\label{WF}
 |\Psi({\bf r},t)\rangle = \sum_n \sum_{\bf K} a_{n{\bf K}}(t) \tilde{\Psi}_{n{\bf K}}({\bf r},t)
 e^{-\frac{i}{\hbar} \int_{t_0}^t \varepsilon_{n}({\bf k})d\tau}.
\end{equation}
Putting Eq.~(\ref{WF}) into Eq.~(\ref{sHr}) and using the orthogonality of $\tilde{\Psi}_{n{\bf K}}$ noting that $(\tilde{\Psi}_{n{\bf K}}, \tilde{\Psi}_{n^\prime{\bf K}^\prime}) = \delta_{n,n^\prime} \delta_{{\bf K},{\bf K}^\prime}$, we find
\begin{subequations}
\begin{equation}\label{ank}
 \frac{\partial}{\partial t}a_{n{\bf K}}(t) = i\sum_{n^{\prime}} \frac{\partial P_{nn^\prime}(t)}{\partial t} a_{n^\prime{\bf K}}(t),
\end{equation}
where the sum is over all crystal bands at the same ${\bf K}$ in the Brillouin zone (BZ) and
\begin{equation}\label{ddtpnc}
 \frac{\partial P_{nn^\prime}(t)}{\partial t} =  i(\tilde{\Psi}_{n{\bf K}}, \frac{\partial \tilde{\Psi}_{n^\prime{\bf K}}}{\partial t}) e^{\frac{i}{\hbar} \int_{t_0}^{t} [\varepsilon_n ({\bf k}(\tau)) - \varepsilon_{n^{\prime}} ({\bf k}(\tau))] d\tau},
\end{equation}
the probability transition rate. In making use of the instantaneous eigenstates as noted in Eq.~(\ref{psin1}), and after calculating $(\tilde{\Psi}_{n{\bf K}}, \frac{\partial \tilde{\Psi}_{n^\prime{\bf K}}}{\partial t})$ with $\hbar \dot{\bf k}(t) = {\bf F}(t)$,
\begin{equation}\label{ddtpnc1}
 (\tilde{\Psi}_{n{\bf K}}, \frac{\partial \tilde{\Psi}_{n^\prime{\bf K}}}{\partial t}) =
 \frac{1}{i\hbar} {\bf F}(t) \cdot {\bf R}_{nn^\prime}({\bf k}(t)),
\end{equation}
\end{subequations}
we find the probability transition
\begin{subequations}
\begin{eqnarray}\label{pntd}
 P_{nn^\prime}(t) = \frac{1}{\hbar} \int_{t_0}^t dt^\prime {\bf F}(t^\prime) \cdot {\bf R}_{nn^\prime}({\bf k}(t^\prime)) \nonumber \\
 \times e^{\frac{i}{\hbar} \int_{t_0}^{t^\prime} [\varepsilon_n ({\bf k}(\tau)) - \varepsilon_{n^{\prime}} ({\bf k}(\tau))] d\tau},
\end{eqnarray}
where
\begin{equation}\label{Rnn}
  {\bf R}_{nn^\prime}({\bf k}(t)) = \frac{i}{\Omega_c} \int_{\Omega_c} d{\bf r} u^{\ast}_{n{\bf k}(t)}({\bf r}) {\bf \nabla}_{\bf k} u_{n^\prime{\bf k}(t)}({\bf r}),
\end{equation}
\end{subequations}
$\Omega_c$ is the volume of the unit cell, ${\bf R}_{nn^{\prime}}({\bf k}) = {\bf R}^{\ast}_{n^{\prime}n}({\bf k})$, and $P_{nn^{\prime}}(t) =  P^{\ast}_{n^{\prime}n}(t)$. We note that Eq.~(\ref{ank}) along with Eq.~(\ref{ddtpnc}) are often
referred~[\onlinecite{Zhu2,Hathwar,Linaertent}] to as the Krieger-Iafrate (K-I) equation derived previously~[\onlinecite{Krieger}]; this term appears in an augmented form as
\begin{equation}\label{ank1}
 i\hbar \dot{a}_{n{\bf K}}(t) = \varepsilon_n({\bf k}) a_{n{\bf K}}(t) -
 {\bf F}(t) \cdot \sum_{n^{\prime}} {\bf R}_{nn^\prime}({\bf k}(t)) a_{n^\prime{\bf K}}(t).
\end{equation}
Here, Eq.~(\ref{ank}) and Eq.~(\ref{ank1}) are equivalent, and are related by an integration factor. Yet Eq.~(\ref{ank}) is of central interest when considering electric-field-induced interband transitions~[\onlinecite{Krieger2,Iafrate1}], while Eq.~(\ref{ank1}) is in a form conducive to insertion into high performance computation such as full-band Monte Carlo simulations~[\onlinecite{Zhu2,Hathwar,Linaertent}]. We have historically~[\onlinecite{Krieger2,Iafrate1,JHe}] used the form embraced in Eqs.~(\ref{ank})-(\ref{ddtpnc}) since $P_{nn^\prime}(t)$ of Eq.~(\ref{pntd}) expresses the direct transition probability to make transitions between states $n$ and $n^\prime$ under the influence of the electric field.

Lastly, we note that the phase of $|\Psi({\bf r},t)\rangle$ in Eq.~(\ref{WF}) is typically chosen so that ${\bf R}_{nn}({\bf k}) = 0$, a provision which assumes that a crystal possesses an inversion symmetry. If inversion symmetry is broken, then ${\bf R}_{nn}({\bf k})$ is nonzero and needs to be retained. This gives rise to Berry phase effects~[\onlinecite{Zak}] which are included implicitly in part of the initial condition of $a_{n{\bf K}}(t)$ of Eq.~(\ref{ank1}) as $a_{n{\bf K}}(t) e^{-\frac{i}{\hbar} \int_{t_0}^{t} \bar{\varepsilon}_n ({\bf k}(\tau)) d\tau}$, where $\bar{\varepsilon}_n({\bf k}) = \varepsilon_n({\bf k}) - {\bf F}(t) \cdot {\bf R}_{nn}({\bf k})$. For convenience in the analysis, we consider the situation throughout such that ${\bf R}_{nn} = 0$ so $n^\prime$ is implicitly not equal to $n$ in Eq.~(\ref{ank}) since ${\bf P}_{nn} = 0$.

Central to Bloch dynamics in homogeneous electric fields is the probability transition $ P_{nn^\prime}(t)$ from state $n$ to $n^\prime$ which we found to be given in Eq.~(\ref{pntd}). For further analysis in this work, we specify that ${\bf F}(t) = F_0 \hat{\bf x}$, a constant in the $\hat{\bf x}$ direction ($F_0 = eE_0$), and $t_0 = 0$; then, we find
\begin{equation}\label{pntdx}
 P_{nn^\prime}(t) = \frac{F_0}{\hbar} \int_0^t dt^\prime X_{nn^\prime}(t^\prime) e^{\frac{i}{\hbar} \int_0^{t^\prime} [\varepsilon_n ({\bf k}(\tau)) - \varepsilon_{n^\prime} ({\bf k}(\tau))] d\tau},
\end{equation}
where $X_{nn^\prime} = ({\bf R}_{nn^\prime})_x$. Since $X_{nn^\prime}, \varepsilon_n$, and $\varepsilon_{n^\prime}$ are periodic in ${\bf G}$, the reciprocal lattice vectors, we let $t = N\tau_B$, where $\tau_B =2\pi/\omega_B$ is the Bloch period, $\omega_B = F_0a/\hbar$ is the Bloch frequency, $a$ is the lattice constant in the $\hat{\bf x}$ direction, and $N$ is a positive integer. Then it follows~[\onlinecite{Iafrate1}] that $P_{nn^\prime}(N\tau_B)$ can be summed to give
\begin{subequations}
\begin{equation}\label{pnNtB}
 P_{nn^\prime}(N\tau_B) = \frac{1 - e^{iN\beta_{nn^\prime}}}{1 - e^{i\beta_{nn^\prime}}} P_{nn^\prime}(\tau_B),
\end{equation}
so that
\begin{equation}\label{pnN}
 |P_{nn^\prime}(N\tau_B)|^2 = \frac{\sin^2(\beta_{nn^\prime} N/2)}{\sin^2(\beta_{nn^\prime}/2)} |P_{nn^\prime}(\tau_B)|^2,
\end{equation}
where
\begin{equation}\label{betan}
 \beta_{nn^\prime} = \frac{1}{F_0} \int_{-G/2}^{G/2} [\varepsilon_n ({\bf k}) - \varepsilon_{n^\prime} ({\bf k})] dk_x,
\end{equation}
and
\begin{eqnarray}\label{pntauB}
 P_{nn^\prime}(\tau_B) = \frac{F_0}{\hbar}\int_0^{\tau_B} dt^\prime X_{nn^\prime}(t^\prime) \nonumber \\
 \times e^{\frac{i}{\hbar} \int_0^{t^\prime} [\varepsilon_n ({\bf k}(\tau)) - \varepsilon_{n^\prime} ({\bf k}(\tau))] d\tau}.
\end{eqnarray}
Through the transformation to ${\bf k}$-space using $k_x(t) = \hbar^{-1}F_0t + K_x$, for $K_x = 0$ and $P_{nn^\prime}(\tau_B) \rightarrow P_{nn^\prime}(G)$ ($G = G_x = 2\pi/a$), we get
\begin{equation}\label{pnG}
 P_{nn^\prime}(G) = \int_{-G/2}^{G/2} dk_x X_{nn^\prime}(k_x) e^{\frac{i}{F_0} \int_0^{k_x} [\varepsilon_n ({\bf k}^\prime) - \varepsilon_{n^\prime} ({\bf k}^\prime)] dk^\prime_x}.
\end{equation}
\end{subequations}
Therefore
\begin{subequations}
\begin{equation}\label{pnNL}
 |P_{nn^\prime}(N\tau_B)|^2 = \frac{\sin^2(\beta_{nn^\prime} N/2)}{\sin^2(\beta_{nn^\prime}/2)} |P_{nn^\prime}(G)|^2.
\end{equation}
Importantly, for $N$ large ($N \gg 1$), that is for long times ($t \gg \tau_B$), Eq.~(\ref{pnNL}) becomes
\begin{equation}\label{pnNL/N}
 \frac{|P_{nn^\prime}(N\tau_B)|}{N} =  |P_{nn^\prime}(G)|.
\end{equation}
\end{subequations}
Thus, the probability transition (a dimensionless quantity) for one cycle over BZ is the probability ``per unit time", and is therefore key to the dynamics of the periodic system under consideration.  It is interesting to note from Eq.~(\ref{pnN}) that maximum growth of the probability transition from states $n$ to $n^\prime$ occurs after many cycles of $\tau_B$, that is, as $N$ becomes large. As $N$ gets large in Eq.~(\ref{pnN}), the $N$-dependent term limits as $N^2$ for $N$ large with shrinking width $1/N$, provided $\beta_{n,n^\prime} \rightarrow 2\pi(m + \delta)$, where $m$ is a positive integer and $\delta$ is vanishingly small. Thus, $\beta_{n,n^\prime} = 2\pi m$ serves as {\it a selection rule} for $n \rightarrow n^\prime$ transitions, affirming the Wannier-Stark quantization rule~[\onlinecite{Krieger}].

\section{Bloch electron time evolution analysis}\label{Section 3}

In this Section, the time evolution for the Bloch electron is determined by analyzing the Schr\"{o}dinger equation of Eq.~(\ref{sHr}) in sequential time ranges relevant to the overall Bloch motion. In particularly, we focus on the three time spans in the analysis: 1) The time $t = 0^{+}$ just after initial injection into an energy band; 2) For initial $K_x = - \pi/a$, while noting from Eq.~(\ref{Not1}) that ${\bf k}(t) = {\bf K}_0 + {\bf k}_c(t)$, we choose the time, $t$, such that $({\bf k}_c(t))_x \equiv k_c(t) \ll \pi/a$, thereby implying that the electron has moved only a small fraction of the Brillouin zone from initial injection under the action of the field, i.e., $(F_0t/\hbar) \ll \pi/a$, for a constant field, $F_0$; 3) Hereafter, the time of transport is $N\tau_B$, where $\tau_B$ is the Bloch period and $N$ is a positive integer.

For each time span, we determine the appropriate wave function from the time-dependent Schr\"{o}dinger equation, and then calculate the expectation value of the momentum operator (and, when appropriate, its time derivative) to study the inertial resistance to acceleration as a determinant in evaluating motional behavior.

In expressing the solution to Eq.~(\ref{sHr}) in terms of the complete set of states from Eq.~(\ref{psin1}), we get Eq.~(\ref{WF}) where the coefficients, $a_{n{\bf K}}(t)$, are given by Eqs.~(\ref{psin1}) and (\ref{pntd}) as
\begin{subequations}
\begin{eqnarray}\label{ank2}
  a_{n{\bf K}}(t) - a_{n{\bf K}}(0) = \frac{i}{\hbar} \int_0^t dt^\prime \sum_{n^\prime}{\bf F}(t^\prime) \cdot {\bf R}_{nn^\prime}({\bf k}(t^\prime)) \nonumber \\
  \times e^{\frac{i}{\hbar} \int_0^{t^\prime} [\varepsilon_n ({\bf k}(\tau)) - \varepsilon_{n^{\prime}} ({\bf k}(\tau))] d\tau} a_{n^\prime{\bf K}}(t^\prime).   \;\;\;\;\;
\end{eqnarray}
For all three temporal spans to be considered (i.e., 1, 2, and 3 above), we apply Eq.~(\ref{ank2}) under the same initial condition
\begin{equation}\label{incond}
 a_{n{\bf K}}(t_0) = \delta_{n,n_0} \delta_{{\bf K},{\bf K}_0}.
\end{equation}
\end{subequations}
Thus, in Eq.~(\ref{ank2}), we insert the initial condition (\ref{incond}) into the right-hand side under the integral for $a_{n^\prime{\bf K}}(t^\prime)$, so that
\begin{eqnarray}\label{anK0}
  a_{n{\bf K}_0}(t) - \delta_{n,n_0} = \frac{i}{\hbar} \int_0^t dt^\prime {\bf F}(t^\prime) \cdot {\bf R}_{nn_0}({\bf k}(t^\prime)) \nonumber  \\
  \times e^{\frac{i}{\hbar} \int_0^{t^\prime} [\varepsilon_n ({\bf k}(\tau)) - \varepsilon_{n_0} ({\bf k}(\tau))] d\tau};
\end{eqnarray}
$a_{n({\bf K} \neq{\bf K}_0)}(t) = 0$, with ${\bf k}(t^\prime) = {\bf K}_0 + {\bf k}_c(t^\prime)$, and ${\bf k}(\tau) = {\bf K}_0 + {\bf k}_c(\tau)$.

We now analyze the temporal situations for time spans defined by 1, 2, and 3 using Eq.~(\ref{anK0}), the so-called {\it early time solution} to Eq.~(\ref{ank2}), that is, Eq.~(\ref{anK0}), obtained by replacing the coefficient $a_{n^\prime{\bf K}}(t^\prime)$ under the integral of Eq(20) by its initial value at $t = t_0$.

{\it Early time domain.}
In this case, we consider the time just after injection, $t = t_0 + 0^{+}$, that is, $0^{+} \equiv t - t_0 = \delta t \ll \hbar/|\varepsilon_n - \varepsilon_{n_0}|$. Then, the integral of Eq.~(\ref{anK0}) can be estimated as
\begin{equation}\label{Integral}
 a_{n{\bf K}_0}(t) - a_{n{\bf K}_0}(0) \simeq \frac{i}{\hbar} {\bf F}_0 \cdot {\bf R}_{nn_0}({\bf K}_0) \delta t.
\end{equation}
The wave function of Schr\"{o}dinger equation in Eq.~(\ref{WF}) is obtained at $t = t_0 + 0^{+}$ as
\begin{equation}\label{WF1}
 |\Psi({\bf r},\delta t)\rangle \simeq \tilde{\Psi}_{n_0{\bf K}_0}({\bf r})  + \sum_{n \neq n_0} \frac{i}{\hbar} {\bf F}_0 \cdot {\bf R}_{nn_0}({\bf K}_0) \tilde{\Psi}_{n{\bf K}_0}({\bf r}) \delta t.
\end{equation}

In calculating the velocity, ${\bf v}(\delta t)$, to order $\delta t$, using the wave function of Eq.~(\ref{WF1}) and noting that $\hat{\bf v} = \hat{\bf p}/m_0$, we get
\begin{equation}\label{Veloc1}
 {\bf v}(\delta t) = \langle \Psi({\bf r},\delta t)|\frac{\hat{\bf p}}{m_0}|\Psi({\bf r},\delta t) \rangle;
 \nonumber
\end{equation}
and using Eq.~(\ref{WF1}), to order $\delta t$, we find
\begin{eqnarray}\label{Veloc2}
 {\bf v}(\delta t) = \frac{1}{m_0}{\bf p}_{n_0{\bf K}_0n_0{\bf K}_0}  \nonumber \\
 + \frac{i}{\hbar m_0} \delta t {\bf F}_0 \cdot \sum_{n\neq n_0}\Big[{\bf R}_{n_0 n}({\bf K}_0){\bf p}_{n{\bf K}_0 n_0{\bf K}_0}  \nonumber \\
 - {\bf R}_{nn_0}({\bf K}_0){\bf p}_{n_0{\bf K}_0 n{\bf K}_0}\Big].
\end{eqnarray}
In Eq.~(\ref{Veloc2}), ${\bf F}_0$ is taken in $``$dot product" with ${\bf R}_{n_0 n}$ and ${\bf R}_{nn_0}$. For ${\bf F}_0 = F_0 \hat{\bf x}$, and noting that
\begin{subequations}
\begin{equation}\label{R}
 {\bf R}_{n^\prime n}({\bf K}_0) = \frac{i\hbar}{m_0} \frac{{\bf p}_{n^\prime{\bf K}_0 n{\bf K}_0}}{\varepsilon_{n^\prime}({\bf K}_0) - \varepsilon_{n} ({\bf K}_0)},
\end{equation}
the velocity $x$-component, $v_x(\delta t)$, of Eq.~(\ref{Veloc2}) can be expressed as
\begin{eqnarray}\label{Veloc3}
 v_x(\delta t) = v_{n_0}({\bf K}_0 + \hbar^{-1}{\bf F}_0 \delta t)  \nonumber  \\
 + 2\delta t \frac{F_0}{m_0^2} \sum_{n\neq n_0} \frac{({\bf p}_{n{\bf K}_0n_0{\bf K}_0})_x ({\bf p}_{n_0{\bf K}_0 n{\bf K}_0})_x}{\varepsilon_n({\bf K}_0) - \varepsilon_{n_0}({\bf K}_0)}.
\end{eqnarray}
\end{subequations}
Using the $f$-sum rule which states
\begin{subequations}
\begin{equation}\label{f-sum}
 \frac{2}{m_0} \sum_{n \neq n_0} \frac{|{\bf p}_{n{\bf K}n_0{\bf K}}|^2}{\varepsilon_n({\bf K}) - \varepsilon_{n_0}({\bf K})}
 = 1 - \frac{m_0}{\hbar^2} \frac{\partial^2 \varepsilon_{n_0}({\bf K})}{\partial {\bf K}^2},
\end{equation}
the second term on the right-hand side of Eq.~(\ref{Veloc3}) reduces to
\begin{equation}\label{Veloc4}
 2\delta t \frac{F_0}{m_0^2} \sum_{n\neq n_0} \frac{|\big({\bf p}_{n{\bf K}_0 n_0{\bf K}_0}\big)_x|^2}{\varepsilon_n({\bf K}_0) - \varepsilon_{n_0}({\bf K}_0)} = F_0 \delta t \Big(\frac{1}{m_0} - \frac{1}{m^{\ast}}\Big);
\end{equation}
\end{subequations}
here, $m^{\ast}$ is given by
\begin{subequations}
\begin{equation}\label{EffMass}
 \frac{1}{m^{\ast}} = \frac{1}{\hbar} {\bf \nabla}_{\bf K} \cdot {\bf v}_{n_0}({\bf K})\Big|_{{\bf K}={\bf K}_0} = \frac{1}{\hbar^2} {\bf \nabla}^2_{\bf K} \varepsilon_{n_0}({\bf K}) \Big|_{{\bf K}={\bf K}_0},
\end{equation}
the well-known effective mass. We obtain $v_x(\delta t)$ from Eq.~(\ref{Veloc3}) to order ${\cal O}(\delta t)$ as
\begin{eqnarray}\label{Veloc5}
 v_x(\delta t) \simeq v_{n_0}({\bf K}_0) + \frac{\delta t}{\hbar} {\bf F}_0 \cdot {\bf v}_{n_0}({\bf K})\Big|_{{\bf K}={\bf K}_0}  \nonumber \\
 + F_0 \delta t \Big(\frac{1}{m_0} - \frac{1}{m^{\ast}}\Big) = v_{n_0}({\bf K}_0) + \frac{F_0}{m_0}\delta t.
\end{eqnarray}
\end{subequations}
Thus, at time $\delta t = 0^{+}$, the momentum proceeds by the free mass inertia in the earliest moment after injection. The result of Eq.~(\ref{Veloc5}) was obtained by Adams and Argyres~[\onlinecite{Adams}].

{\it Intermediate time domain.}
Next, for the initial wave-vector condition of ${\bf K}_0 = - \frac{\pi}{a}\hat{\bf x} = -\frac{G}{2}\hat{\bf x}$, we consider the time letting ${\bf k}(t) = {\bf K}_0 + {\bf k}_c(t)$, where ${\bf k}_c(t)$ given in Eq.~(\ref{Not2}) is such that $k_c(t) \ll \frac{G}{2}$. Then, in the special case where ${\bf F}(t) = F_0 \hat{\bf x}$, $F_0$ is a constant, we obtain ${\bf k}_c(t) = \frac{F_0}{\hbar}t \hat{\bf x}$. Then, it follows that $t \ll (\pi\hbar/aF_0) = \tau_B/2$, where $\tau_B$ is the Bloch period. Thus, in this scenario,
the electron displaces from the initial ${\bf K}_0$-value by a small fraction of the Brillouin zone. In this case, ${\bf k}(t) \simeq {\bf K}_0$, a constant. In particular, considering ${\bf F}(t) = {\bf F}_0$, a constant, then $a_{n{\bf K}}(t)$ of Eq.~(\ref{anK0}), after integration, becomes
\begin{eqnarray}\label{anK2}
  a_{n{\bf K}}(t) = a_{n{\bf K}}(0)  \nonumber  \\
  - {\bf F}_0 \cdot {\bf R}_{nn_0}({\bf K_0})
  \frac{e^{\frac{i}{\hbar} \int_0^t [\varepsilon_n({\bf K}_0) - \varepsilon_{n_0}({\bf K}_0)]d\tau} - 1}{\varepsilon_n ({\bf K}_0) - \varepsilon_{n_0}({\bf K}_0)};
\end{eqnarray}
here, $a_{n{\bf K}}(0) = \delta_{nn_0} \delta_{{\bf K}{\bf K}_0}$, as noted previously. Then, the wave function of Eq.~(\ref{WF}) becomes
\begin{eqnarray}\label{WF2}
  |\Psi({\bf r},t)\rangle \simeq \tilde{\Psi}_{n_0{\bf K}_0}({\bf r})e^{-\frac{i}{\hbar} \int_0^t \varepsilon_{n_0}({\bf k})d\tau}  \nonumber  \\
  - \sum_{n\neq n_0} {\bf F}_0 \cdot {\bf R}_{nn_0}({\bf K}_0)
  \frac{e^{\frac{i}{\hbar} \int_0^t [\varepsilon_n({\bf K}_0) - \varepsilon_{n_0}({\bf K}_0)]d\tau} - 1}{\varepsilon_n ({\bf K}_0) - \varepsilon_{n_0}({\bf K_0})}  \nonumber  \\
  \times e^{-\frac{i}{\hbar} \int_0^t \varepsilon_n({\bf k})d\tau} \tilde{\Psi}_{n{\bf K}_0}({\bf r}). \;\;\;
\end{eqnarray}

Using $|\Psi({\bf r},t)\rangle$ of Eq.~(\ref{WF2}) with $\tilde{\Psi}_{n{\bf K}}$ of Eq.~(\ref{psin1}), we find the velocity for ${\bf F}(t) = F_0 \hat{\bf x}$ to ${\cal O}(F_0)$ as
\begin{eqnarray}\label{velocity6}
  v_x(t) \simeq \frac{1}{\hbar} \frac{\partial \varepsilon_{n_0}({\bf k}(t))}{\partial k_x} +
  F_0 \frac{2\hbar}{m_0^2} \sum_{n \neq n_0} \frac{|{\bf p}_{n{\bf K}_0,n_0{\bf K}_0}|^2}{(\varepsilon_n - \varepsilon_{n_0})^2}  \nonumber  \\
  \sin \Bigl(\frac{1}{\hbar} \int_0^t [\varepsilon_n({\bf k}(\tau)) - \varepsilon_{n_0}({\bf k}(\tau))]d\tau \Bigr).
\end{eqnarray}
It follows from Eq.~(\ref{velocity6}) that the acceleration becomes, using $v_x = p_x/m_0$,
\begin{subequations}
\begin{eqnarray}\label{acceleration}
  \frac{dv_x}{dt} = \frac{1}{\hbar} \frac{\partial^2 \varepsilon_{n_0}({\bf k}(t))}{\partial k^2_x} \dot{k_x}    + F_0\frac{2}{m_0^2} \sum_{n \neq n_0} \frac{|{\bf p}_{n{\bf K}_0,n_0{\bf K}_0}|^2}{\varepsilon_n - \varepsilon_{n_0}} \nonumber  \\
  \cos \Bigl(\frac{1}{\hbar} \int_0^t [\varepsilon_n({\bf k}(\tau)) - \varepsilon_{n_0}({\bf k}(\tau))]d\tau \Bigr);
  \;\;\;\;
\end{eqnarray}
and since $F_0 = \hbar \dot{k_x}$, Eq.~(\ref{acceleration}) can be written as
\begin{eqnarray}\label{acceleration1}
  \frac{dv_x}{dt} = \frac{1}{\hbar^2} \frac{\partial^2 \varepsilon_{n_0}({\bf k}(t))}{\partial k^2_x} F_0    + F_0\frac{2}{m_0^2} \sum_{n \neq n_0} \frac{|{\bf p}_{n{\bf K}_0n_0{\bf K}_0}|^2}{\varepsilon_n - \varepsilon_{n_0}}
  \nonumber  \\
  \cos \Bigl(\frac{1}{\hbar} \int_0^t [\varepsilon_n({\bf k}(\tau)) - \varepsilon_{n_0}({\bf k}(\tau))]d\tau \Bigr).
  \;\;\;\;\;
\end{eqnarray}
\end{subequations}
We note that $v_x(t)$ of Eq.~(\ref{velocity6}) contains the usual $n_0$-band average velocity term plus an additional oscillatory time-dependent contribution from all bands beyond the initial band, $n_0$. The oscillatory term carries over into the acceleration term of Eq.~(\ref{acceleration1}) as well.

In considering the temporal limits of the oscillatory term in Eq.~(\ref{velocity6}), we note that as $t \rightarrow 0^{+}$, Eq.~(\ref{velocity6}),
\begin{equation}\label{Limit}
 \frac{\sin \Bigl(\frac{1}{\hbar} \int_0^t [\varepsilon_n({\bf k}(\tau)) - \varepsilon_{n_0}({\bf k}(\tau))]d\tau \Bigr)}{\varepsilon_n({\bf k}(\tau)) - \varepsilon_{n_0}({\bf k}(\tau))} \rightarrow \frac{t}{\hbar};
\end{equation}
Then for $t \rightarrow 0^{+}$, Eq.~(\ref{velocity6}) becomes
\begin{subequations}
\begin{equation}\label{velocity6-1}
  v_x(t) \simeq \frac{1}{\hbar} \frac{\partial \varepsilon_{n_0}(k_x(t))}{\partial k_x} +
  F_0\frac{2}{m_0^2} t\sum_{n \neq n_0} \frac{|{\bf p}_{n{\bf K}_0n_0{\bf K}_0}|^2}{\varepsilon_n - \varepsilon_{n_0}}.
\end{equation}
Taking into account the $f$-sum rule which states
\begin{equation}\label{f-sum}
 \frac{2}{m_0} \sum_{n \neq n_0} \frac{|{\bf p}_{n{\bf K}_0n_0{\bf K}_0}|^2}{\varepsilon_n - \varepsilon_{n_0}}
 = 1 - \frac{m_0}{\hbar^2} \frac{\partial^2 \varepsilon_{n_0}(k_x(t))}{\partial k^2_x},  \nonumber
\end{equation}
we can express $v_x(t)$ of Eq.~(\ref{velocity6-1}) as
\begin{equation}\label{velocity6-2}
  v_x(t) \simeq \frac{1}{\hbar} \frac{\partial \varepsilon_{n_0}(k_x(t))}{\partial k_x} +
  F_0t\frac{1}{m_0} \Big[1 - \frac{m_0}{\hbar^2} \frac{\partial^2 \varepsilon_{n_0}(k_x(t))}{\partial k^2_x}\Big].
\end{equation}
\end{subequations}
And since $k_x(t) = K_0 + F_0t/\hbar$, we then find to first order in $F_0$ that
\begin{eqnarray}\label{velocity6-3}
  \frac{1}{\hbar} \frac{\partial \varepsilon_{n_0}(k_x(t))}{\partial k_x} =  v_{n_0}(K_0 + \frac{F_0}{\hbar}t)   \nonumber  \\
  \simeq v_{n_0}(K_0) + \frac{1}{\hbar^2} \frac{\partial^2 \varepsilon_{n_0}(k_x)}{\partial k^2_x} F_0t.   \nonumber
\end{eqnarray}
Thus, it follows that
\begin{equation}\label{velocity6-4}
  v_x(t) = v_{n_0}({\bf K}_0) + \frac{F_0}{m_0}t
\end{equation}
and
\begin{equation}\label{velocity6-5}
  \frac{dv_x(t)}{dt} = \frac{F_0}{m_0},
\end{equation}
that is, real mass acceleration at $t \rightarrow 0^{+}$ as noted in time span 1.

In Eq.~(\ref{acceleration1}), when $t$ gets large, that is, theoretically $t \rightarrow \infty$, the cosine terms
oscillate out of phase with each other and thus sum to zero, so that the final term becomes
\begin{equation}\label{acceleration2}
  \frac{dv_x}{dt} = \frac{1}{\hbar^2} \frac{\partial^2 \varepsilon_{n_0}({\bf k}(t))}{\partial k^2_x} F_0,
\end{equation}
the effective mass result for electron acceleration.

In estimating the time that must elapse before the effective mass is apparent, we note, in Eq.~(\ref{acceleration1}), that the time, $\triangle t$, necessary for the most slowly varying term to go through one cycle is
\begin{subequations}
\begin{equation}\label{Esitmat1}
 \triangle t = \frac{2\pi\hbar}{\varepsilon_{n_0 + 1} - \varepsilon_{n_0}},
\end{equation}
where $\varepsilon_{n_0 + 1} > \varepsilon_{n_0}$ is the nearest band to $\varepsilon_{n_0}$. Hence, the estimation suggests
\begin{equation}\label{Esitmat2}
 \triangle t \simeq \frac{2\pi\hbar}{E_g} = \frac{4.17}{E_g(eV)}\times 10^{-15} s,
\end{equation}
\end{subequations}
where $E_g$ is the energy gap from $\varepsilon_{n_0}$ to $\varepsilon_{n_0 + 1}$. It is noted that for bulk semiconductors with $E_g \sim 1$ eV, the order of magnitude of $\triangle t$ is in the femtosecond range; whereas for narrow band-gap semiconductors ($E_g \sim 10^{-1} eV$) and superlattice materials with $E_g \sim 10^{-2}$ eV, this increases to $10^{-13} s$ range. Thus, when in high speed femtosecond-attosecond time-resolved spectroscopy, the Bloch dynamics for materials with such $E_g$ values is apt to be in dynamical non-equilibrium, described by the transient process from the bare mass to the effective mass dynamics.

We note that the oscillatory velocity term in Eq.~(\ref{velocity6}) can be viewed as a field-dependent generalization of the natural {\it Zitterbewegung} (ZB)~[\onlinecite{Schrodinger}]-like presence discussed by Zawadzki and Rusin~[\onlinecite{Zawadzki}] for semiconductors and others~[\onlinecite{David,Cao}]. Although the classical ZB-like phenomena are most pronounced in two-band Dirac-like systems, in multiband solid-state systems, non-negligible coupling between the bands gives rise to necessary “sum over states”, which over time, probabilistically smears out in a time average, the two-band model effects, thus resulting in an effective mass picture. We estimated in Eq.~(\ref{Esitmat1}) the time duration required for the effective mass to be apparent after initial injection. This time duration is synonymous with the time estimate for the ZB-like oscillations to be observable.

{\it Time domain of multiple Bloch periods.}
Lastly, we consider the long-time regime defined by $t = N\tau_B$ where $N$ is a positive integer, and for a constant electric field, ${\bf F} = F_0 \hat{\bf x}$, with $\tau_B = 2\pi\hbar/aF_0$. For this particular scenario, we consider the solution to Eq.~(\ref{ank}) for $t = N\tau_B$. In general, the exact solution to Eq.~(\ref{ank}) is not tractable. But an excellent approximate solution is achieved by using the Wigner-Weisskopf approximation (WWA). The WWA proceeds as follows. For $a_{n^\prime{\bf K}}(t)$ on the right-hand side of Eq.~(\ref{ank}), we assume that
\begin{equation}\label{ank3}
 \frac{\partial}{\partial t}a_{n^\prime{\bf K}}(t) = i\frac{\partial P_{n^\prime n}(t)}{\partial t} \, a_{n{\bf K}}(t).
\end{equation}
This approximation directly backfills all the states $n^\prime$ to the state $n$ while ignoring multiple reflections. Thus, in integrating the equation for $a_{n^\prime{\bf K}}(t)$ in Eq.~(\ref{ank3}), and inserting it in Eq.~(\ref{ank}), we get
\begin{equation}\label{ank4}
 \frac{\partial}{\partial t}a_{n{\bf K}}(t) = -\sum_{n^\prime} \frac{\partial P_{nn^\prime}(t)}{\partial t} \int_{t_0}^t \frac{\partial P^{\ast}_{nn^\prime}(t^\prime)}{\partial t^\prime} \, a_{n{\bf K}}(t^\prime)dt^\prime,
\end{equation}
where we have used that $P_{n^\prime n}(t^\prime) = P^{\ast}_{nn^\prime}(t^\prime)$. Here, $a_{n^\prime{\bf K}}(t_0) = 0$ so that the states $a_{n^\prime{\bf K}}(t)$ are available for occupation. Integrating by parts on the right-hand side, we get
\begin{eqnarray}\label{ank5}
 \frac{\partial}{\partial t} a_{n{\bf K}}(t) = -\sum_{n^\prime} \frac{\partial P_{nn^\prime}(t)}{\partial t} \Big\{P^{\ast}_{nn^\prime}(t) \, a_{n{\bf K}}(t)  \nonumber \\
 - \int_{t_0}^t P^{\ast}_{nn^\prime}(t^\prime) \frac{\partial a_{n{\bf K}}(t^\prime)}{\partial t^\prime} dt^\prime \Big\}    \nonumber  \\
 = -\sum_{n^\prime} \frac{\partial P_{nn^\prime}(t)}{\partial t} P^{\ast}_{nn^\prime}(t) \, a_{n{\bf K}}(t) + {\cal O}(P^4).
\end{eqnarray}
Dropping high-order terms in ${\cal O}(P^4)$, since $P_{nn^\prime}(t)$ is exponentially small (Zener tunneling magnitude), we have
\begin{equation}\label{ank6}
 \frac{\partial}{\partial t} a_{n{\bf K}}(t)  = -\Big(\sum_{n^\prime} \frac{\partial P_{nn^\prime}(t)}{\partial t} P^{\ast}_{nn^\prime}(t)\Big) \, a_{n{\bf K}}(t).
\end{equation}
Therefore, after integration
\begin{equation}\label{ank7}
 a_{n{\bf K}}(t) = a_{n{\bf K}}(t_0) \, e^{-\sum_{n^\prime} \int_{t_0}^t dt^\prime  \frac{\partial P_{nn^\prime}(t^\prime)}{\partial t^\prime} P^{\ast}_{n n^\prime}(t^\prime)},
\end{equation}
so that
\begin{equation}\label{anksquar}
 |a_{n{\bf K}}(t)|^2 = |a_{n{\bf K}}(t_0)|^2 e^{-\sum_{n^\prime} |P_{nn^\prime}(t)|^2}.
\end{equation}

Now, using $|\Psi({\bf r},t)\rangle$ of Eq.~(\ref{WF}), we calculate
\begin{eqnarray}\label{v(t)}
 {\bf v}(t) = \langle \Psi({\bf r},t)|\frac{\hat{\bf p}}{m_0}|\Psi({\bf r},t) \rangle \nonumber  \\
 = \sum_{n,n^\prime} \sum_{{\bf K},{\bf K}^\prime} a^{\ast}_{n^\prime{\bf K}^\prime}(t) a_{n{\bf K}}(t)
 \langle \tilde{\Psi}_{n^\prime {\bf K}^\prime}|\frac{\hat{\bf p}}{m_0}|\tilde{\Psi}_{n {\bf K}} \rangle \nonumber \\
 \times e^{\frac{i}{\hbar} \int_{t_0}^t [\varepsilon_{n^\prime}({\bf k}^\prime) - \varepsilon_{n}({\bf k})]d\tau},
\end{eqnarray}
where
\begin{eqnarray}\label{v(t)1}
\langle \tilde{\Psi}_{n^\prime {\bf K}^\prime}|\frac{\hat{\bf p}}{m_0}|\tilde{\Psi}_{n {\bf K}} \rangle \nonumber \\
= \left\{ \begin{array}{ll}
\frac{i}{\hbar} [\varepsilon_{n^\prime}({\bf k}) - \varepsilon_{n}({\bf k})]{\bf R}_{n^\prime n}({\bf k})\delta_{{\bf K},{\bf K}^\prime}, & n \neq n^\prime ;\\
\frac{1}{\hbar} {\bf \nabla}_{\bf k}\varepsilon_{n}({\bf k})\delta_{{\bf K},{\bf K}^\prime}, & n = n^\prime.
\end{array} \right.
\end{eqnarray}

In forming $a_{n{\bf K}}(t) a^{\ast}_{n^\prime{\bf K}^\prime}(t)$, we use Eq.~(\ref{ank7}). Thus, to lowest order retaining terms in $n, n^\prime$ only, we obtain
\begin{eqnarray}\label{ankank}
 a^{\ast}_{n^\prime{\bf K}^\prime}(t) a_{n{\bf K}}(t) \simeq a^{\ast}_{n^\prime{\bf K}^\prime}(t_0) a_{n{\bf K}}(t_0)
 \nonumber  \\
 \times e^{- 2\int_{t_0}^t dt^\prime  \frac{\partial P_{nn^\prime}(t^\prime)}{\partial t^\prime} P^{\ast}_{n n^\prime}(t^\prime)}.
\end{eqnarray}

We note that the integral term in the exponential of Eq.~(\ref{ankank}) can be written as (with $t_0 = 0$)
\begin{subequations}
\begin{equation}\label{ankankA}
 I(t) = \int_0^t dt^\prime  \frac{\partial P_{nn^\prime}(t^\prime)}{\partial t^\prime} P^{\ast}_{n n^\prime}(t^\prime) = \int_0^{P_{nn^\prime}(t)}  P^{\ast}_{nn^\prime} dP_{nn^\prime}.
\end{equation}
Formally taking $P_{nn^\prime}$ as a complex variable,
\begin{equation}\label{ankankC}
 P_{nn^\prime} = \varrho e^{i\vartheta},
\end{equation}
then the integrand of Eq.~(\ref{ankankA}) becomes $P^{\ast}_{nn^\prime} dP_{nn^\prime} = \varrho d\varrho + i\varrho^2 d\vartheta$, so that $I(t)$ of Eq.~(\ref{ankankA}) is
\begin{equation}\label{ankankE}
 I(t) = \frac{\varrho^2}{2}(1 + 2i\vartheta).
\end{equation}
\end{subequations}
Using $P_{nn^\prime}$ of Eq.~(\ref{ankankC}) in Eq.~(\ref{ankankE}), we find that
\begin{equation}\label{i(t)}
 I(t) = \frac{|P_{nn^\prime}|^2}{2}\big(1 + 2\ln \frac{P_{nn^\prime}}{|P_{nn^\prime}|}\big).
\end{equation}
Thus, Eq.~(\ref{ankank}) can be written as
\begin{eqnarray}\label{ankank1}
 a^{\ast}_{n^\prime{\bf K}^\prime}(t) a_{n{\bf K}}(t) \simeq a^{\ast}_{n^\prime{\bf K}^\prime}(t_0) a_{n{\bf K}}(t_0)
 \nonumber  \\
 \times e^{- |P_{nn^\prime}(t)|^2 \big(1 + 2\ln \frac{P_{nn^\prime}(t)}{|P_{nn^\prime}(t)|}\big)}.
\end{eqnarray}

The equation (\ref{v(t)}) can be written as
\begin{eqnarray}\label{v1}
 {\bf v}(t) = \sum_{n,{\bf K}} |a_{n{\bf K}}(t)|^2 \frac{1}{\hbar} {\bf \nabla}_{\bf k}\varepsilon_{n}({\bf k})
 \nonumber  \\
 + \frac{i}{\hbar} \sum_{n,n^\prime \neq n} \sum_{{\bf K}} a^{\ast}_{n^\prime {\bf K}}(t) a_{n{\bf K}}(t)
 [\varepsilon_{n^\prime}({\bf k}) - \varepsilon_n({\bf k})]  \nonumber \\
 \times {\bf R}_{n^\prime n}({\bf k})
 e^{\frac{i}{\hbar} \int_{t_0}^t [\varepsilon_{n^\prime}({\bf k}) - \varepsilon_{n}({\bf k})]d\tau}.
\end{eqnarray}
In Eq.~(\ref{v1}), $|a_{n{\bf K}}(t)|^2$ and $a^{\ast}_{n^\prime {\bf K}}(t) a_{n{\bf K}}(t)$ are expressed
in terms of $P_{nn^\prime}(t)$ in Eqs.~(\ref{anksquar}) and (\ref{ankank}). $P_{nn^\prime}(t)$ is fully characterized at $t = N\tau_B$ from Eqs.~(\ref{pntdx}), (\ref{pnNtB})-(\ref{pnG}).

\section{Summary and conclusions}\label{Summary}
The temporal evolution for a Bloch electron accelerating in a time-dependent homogeneous electric field is analyzed. Temporal stages are studied from injection through various stages of key evolution. The time-dependent Schr\"{o}dinger equation is analyzed, and expectation values for momentum are evaluated to determine early-time and long-time behavior in a constant, first-order electric field. Results show that the resistance to acceleration at early time differs markedly from the long-time limit due to the dynamics of {\it effective mass} time dependence or dynamical inertial due to the dressing-up of the real mass. In strong electric fields over short time scales, such dynamical effects should be expected as a matter of {\it inertial breathing}.

Noting that the classical ZB-like phenomena are most pronounced in two-band Dirac-like systems, in multiband solid-state systems, non-negligible coupling between the bands gives rise to necessary ``sum over states", which over time, probabilistically smears out in a time average, the two-band model effects thus resulting in the all-important effective mass picture.

In studying the dynamical inertia through the momentum response to a first-order constant electric field, we estimated that the time required to reach dynamical effective mass equilibrium in a semiconductor of unit energy gap is in the femtosecond range. Thus, we are alerted to the notion that {\it inertial breathing} due to strong electric fields can have a significant influence on the Bloch electron dynamics in the early time evolution after initial injection.

\appendix
\section{The time evolution operator}\label{Appendix}
In the development of time evolution theory~[\onlinecite{Roy}], it is assumed that the quantum state of the system evolves from initial time $t_0$ to time $t > t_0$ by means of a unitary transformation~[\onlinecite{Sakurai}] such that
\begin{equation}\label{Utt0}
 \Psi(t) = \hat{U}(t,t_0)\Psi(t_0);
\end{equation}
here, normalization is preserved through the unitary property
\begin{equation}\label{}
 \hat{U^{\dag}}(t,t_0)\hat{U}(t,t_0) = \hat{I}, \nonumber
\end{equation}
where
\begin{equation}\label{}
 \hat{U}(t_0,t_0) = \hat{I}.   \nonumber
\end{equation}

We further consider Eqs.~(\ref{H}) and (\ref{sHr}), and given Eqs.~(\ref{H1}) and (\ref{sHr}), we find that the unitary operator $\hat{U}$ satisfies the equation
\begin{equation}\label{EqU}
 i\hbar \frac{\partial \hat{U}}{\partial t} = \hat{H}(t) \hat{U}.
\end{equation}
Note that for the specific time-dependent Hamiltonian $\hat{H}(t)$ of Eq.~(\ref{H}), we find that the commutator of $\hat{H}(t)$ and $\hat{H}(t^\prime)$ for $t \neq t^\prime$ is
\begin{equation}\label{Commut}
 [\hat{H}(t), \hat{H}(t^\prime)] = \frac{1}{m_0}\big\{[{\bf p}_c(t) - {\bf p}_c(t^\prime)] \cdot \hat{\bf p}\big\}V_c({\bf r}).
\end{equation}
This indicates that the formal solution to Eq.~(\ref{EqU}) takes the form of an infinite Dyson series~[\onlinecite{Sakurai}]. But in our approach, we seek the behavior of Eq.~(\ref{EqU}) by taking the matrix elements of the $\hat{U}$ operator, using the instantaneous eigenstates of the Hamiltonian $\hat{H}(t)$ of Eq.~(\ref{H}). In this regard, for the instantaneous eigenstates of Eq.~(\ref{H1}), namely
\begin{equation}\label{Basisset}
 \hat{H}(t) |\tilde{\Psi}_{n{\bf K}}\rangle = \varepsilon_n({\bf k(t)})|\tilde{\Psi}_{n{\bf K}}\rangle,
\end{equation}
we have obtained the time-dependent wave vector ${\bf k(t)}$ [Eq.~(\ref{Not1})] and the wave function $\tilde{\Psi}_{n{\bf K}} \equiv |n,{\bf K}\rangle$ given in Eq.~(\ref{psin1}). Here, $|n,{\bf K}\rangle$ is a complete set of normalized basis functions with $(\tilde{\Psi}_{n{\bf K}},\tilde{\Psi}_{n^\prime{\bf K}^\prime}) = \delta_{n,n^\prime} \delta_{{\bf K},{\bf K}^\prime}$. Thus, the matrix elements of Eq.~(\ref{EqU}) become
\begin{equation}\label{EqUME}
 i\hbar \big\langle n^\prime {\bf K}^\prime \big| \frac{\partial \hat{U}}{\partial t} \big|n {\bf K}\big\rangle = \langle n^\prime {\bf K}^\prime|\hat{H} \hat{U}|n {\bf K}\rangle.
\end{equation}
For the matrix elements on the right-hand side of Eq.~(\ref{EqUME}), we can write
\begin{subequations}
\begin{eqnarray}\label{RHS}
 \langle n^\prime {\bf K}^\prime|\hat{H} \hat{U}|n {\bf K}\rangle \nonumber \\
 = \sum_{n^{\prime\prime},{\bf K}^{\prime\prime}} \langle n^\prime {\bf K}^\prime|\hat{H}|n^{\prime\prime} {\bf K}^{\prime\prime}\rangle \langle n^{\prime\prime} {\bf K}^{\prime\prime}|\hat{U}|n {\bf K}\rangle  \nonumber \\
 = \varepsilon_{n^\prime}({\bf k^\prime(t)}) \langle n^\prime {\bf K}^\prime|\hat{U}|n {\bf K}\rangle,
\end{eqnarray}
where we have employed Eq.~(\ref{Basisset}) and orthogonality of the states $|n,{\bf K}\rangle$. In addition, for the $\frac{\partial \hat{U}}{\partial t}$ matrix elements in Eq.~(\ref{EqUME}), we can formally write
\begin{eqnarray}\label{Formal}
 \frac{\partial}{\partial t} \langle n^\prime {\bf K}^\prime|\hat{U}|n {\bf K}\rangle = \big\langle n^\prime {\bf K}^\prime \big|\frac{\partial\hat{U}}{\partial t} \big|n{\bf K}\big\rangle  \nonumber \\
 + \int d{\bf r} \Big[\frac{\partial \tilde{\Psi}^{\ast}_{n^\prime {\bf K}^\prime}}{\partial t}\hat{U}\tilde{\Psi}_{n{\bf K}}  \nonumber \\
 + \tilde{\Psi}^{\ast}_{n^\prime {\bf K}^\prime} \hat{U} \frac{\partial \tilde{\Psi}_{n{\bf K}}}{\partial t}\Big].
\end{eqnarray}
The right-hand term of Eq.~(\ref{Formal}) simplifies further by noting that
\begin{eqnarray}\label{Psink}
 i\hbar\frac{\partial \tilde{\Psi}_{n{\bf K}}({\bf r},t)}{\partial t} = {\bf F}(t)\cdot \sum_{n^{\prime\prime}}{\bf R}_{n^{\prime\prime}n}({\bf k})\tilde{\Psi}_{n^{\prime\prime}{\bf K}}({\bf r},t),  \nonumber \\
 -i\hbar\frac{\partial \tilde{\Psi}^{\ast}_{n^\prime{\bf K}^\prime}({\bf r},t)}{\partial t} = {\bf F}(t)\cdot \sum_{n^{\prime\prime}}{\bf R}^{\ast}_{n^{\prime\prime}n^\prime}({\bf k}^\prime)\tilde{\Psi}^{\ast}_{n^{\prime\prime}{\bf K}^\prime}({\bf r},t), \nonumber
\end{eqnarray}
so that the right-hand term becomes
\begin{eqnarray}\label{RHS1}
 \int d{\bf r} \Big[\cdots \Big] = \frac{1}{i\hbar}{\bf F}(t)\cdot \sum_{n^{\prime\prime}}
 \Big[{\bf R}_{n^{\prime\prime}n}({\bf k}) \langle n^\prime {\bf K}^\prime|\hat{U}|n^{\prime\prime} {\bf K}\rangle
 \nonumber  \\
 - {\bf R}^{\ast}_{n^{\prime\prime}n^\prime}({\bf k}^\prime) \langle n^{\prime\prime} {\bf K}^\prime|\hat{U}|n{\bf K}\rangle \Big].   \;\;\;\;
\end{eqnarray}
\end{subequations}
Then, putting Eqs.~(\ref{EqUME}), (\ref{RHS}), and (\ref{RHS1}) into Eq.~(\ref{Formal}), we find that the matrix equation for $\hat{U}$ in symmetric form, using $\sum_{{\bf K}^{\prime\prime}} \delta_{{\bf K},{\bf K}^{\prime\prime}}$ and $\sum_{{\bf K}^{\prime\prime}}\delta_{{\bf K}^\prime,{\bf K}^{\prime\prime}}$, is
\begin{eqnarray}\label{MasterEq}
 i\hbar\frac{\partial}{\partial t} \langle n^\prime {\bf K}^\prime|\hat{U}|n {\bf K}\rangle = \varepsilon_{n^\prime}({\bf k^\prime(t)}) \langle n^\prime {\bf K}^\prime|\hat{U}|n {\bf K}\rangle \nonumber \\
 + {\bf F}(t)\cdot \sum_{n^{\prime\prime}} \sum_{{\bf K}^{\prime\prime}} \Big[{\bf R}_{n^{\prime\prime}n}({\bf k}^{\prime\prime}) \langle n^\prime{\bf K}^\prime|\hat{U}|n^{\prime\prime} {\bf K}^{\prime\prime}\rangle \delta_{{\bf K},{\bf K}^{\prime\prime}} \nonumber \\
 - {\bf R}_{n^\prime n^{\prime\prime}}({\bf k}^{\prime\prime}) \langle n^{\prime\prime} {\bf K}^{\prime\prime}|\hat{U}|n{\bf K}\rangle \delta_{{\bf K}^{\prime},{\bf K}^{\prime\prime}} \Big].
\end{eqnarray}
We will refer to Eq.~(\ref{MasterEq}) as the overall key master equation for the matrix elements of $\hat{U}$.

In solving for the matrix elements in determinant form, we first note in Eq.~(\ref{MasterEq}) that for $n^\prime {\bf K}^\prime = n{\bf K}$, the {\it exact} diagonal matrix element equation can be expressed in terms of
off-diagonal matrix elements as
\begin{eqnarray}\label{MasterEq1}
 i\hbar\frac{\partial}{\partial t} \langle n{\bf K}|\hat{U}|n {\bf K}\rangle = \varepsilon_{n}({\bf k}) \langle n {\bf K}|\hat{U}|n{\bf K}\rangle \nonumber \\
 + {\bf F}(t)\cdot \sum_{n^{\prime\prime}} \sum_{{\bf K}^{\prime\prime}} \Big[{\bf R}_{n^{\prime\prime}n}({\bf k}^{\prime\prime}) \langle n{\bf K}|\hat{U}|n^{\prime\prime} {\bf K}^{\prime\prime}\rangle \delta_{{\bf K},{\bf K}^{\prime\prime}} \nonumber \\
 - {\bf R}_{n n^{\prime\prime}}({\bf k}^{\prime\prime}) \langle n^{\prime\prime} {\bf K}^{\prime\prime}|\hat{U}|n{\bf K}\rangle \delta_{{\bf K},{\bf K}^{\prime\prime}} \Big].
\end{eqnarray}

In an approximation to Eq.~(\ref{MasterEq}), on the right-hand side, for $\sum_{n^{\prime\prime},{\bf K}^{\prime\prime}}$, we retain only terms for $n{\bf K}$ and $n^{\prime}{\bf K}^{\prime}$, dropping all other terms since they are considered to be of a higher order,
\begin{eqnarray}\label{Sumation}
  \sum_{n^{\prime\prime},{\bf K}^{\prime\prime}}\Big[\cdots \Big] \simeq \Big[{\bf R}_{nn}({\bf k}) - {\bf R}_{n^\prime n^\prime}({\bf k}^\prime)\Big] \langle n^\prime{\bf K}^\prime|\hat{U}|n{\bf K}\rangle \nonumber \\
  + {\bf R}_{n^\prime n}({\bf k}^\prime) \Big(\langle n^\prime {\bf K}^\prime|\hat{U}|n^\prime{\bf K}^\prime\rangle - (\langle n {\bf K}|\hat{U}|n{\bf K}\rangle \Big)   \nonumber.
\end{eqnarray}
Then, ignoring the higher order terms projected out of $(n{\bf K}, n^{\prime}{\bf K}^{\prime})$, we find that Eq.~(\ref{MasterEq}) reduces to
\begin{eqnarray}\label{MasterEq2}
 i\hbar\frac{\partial}{\partial t} \langle n^\prime {\bf K}^\prime|\hat{U}|n {\bf K}\rangle \nonumber \\
 = \Big[\varepsilon_{n^\prime}({\bf k}^\prime) + {\bf F}(t)\cdot \Big({\bf R}_{nn}({\bf k}) - {\bf R}_{n^\prime n^\prime}({\bf k}^\prime)\Big) \Big]  \nonumber  \\
 \times \langle n^\prime{\bf K}^\prime|\hat{U}|n{\bf K}\rangle
 + {\bf F}(t) \cdot {\bf R}_{n^\prime n}({\bf k}^\prime)  \nonumber \\
 \times \Big[\langle n^\prime {\bf K}^\prime|\hat{U}|n^\prime{\bf K}^\prime\rangle - \langle n {\bf K}|\hat{U}|n{\bf K}\rangle \Big].  \;\;\;\;
\end{eqnarray}
Equation ~(\ref{MasterEq2}) gives the off-diagonal elements in terms of diagonal elements of the $\hat{U}$ matrix. As such, we can integrate Eq.~(\ref{MasterEq2}) for $ \langle n^\prime {\bf K}^\prime|\hat{U}|n {\bf K}\rangle$ to obtain
\begin{eqnarray}\label{OffDiag}
 \langle n^\prime {\bf K}^\prime|\hat{U}(t)|n {\bf K}\rangle = \langle n^\prime {\bf K}^\prime|\hat{U}(t_0)|n {\bf K}\rangle  \nonumber \\
 \times e^{-\frac{i}{\hbar} \int_{t_0}^t dt^\prime [\varepsilon_{n^\prime}({\bf k}^\prime(t^\prime)) + {\bf F} (t^\prime)\cdot ({\bf R}_{nn}({\bf k}(t^\prime)) - {\bf R}_{n^\prime n^\prime}({\bf k}^\prime(t^\prime)))]}    \nonumber \\
 + \frac{1}{i\hbar} \int_{t_0}^t dt^\prime {\bf F}(t^\prime)\cdot {\bf R}_{n^\prime n}({\bf k}^\prime(t^\prime))
 \Big[\langle n^\prime{\bf K}^\prime|\hat{U}(t^\prime)|n^\prime{\bf K}^\prime\rangle  \nonumber \\
 - \langle n{\bf K}|\hat{U}(t^\prime)|n{\bf K}\rangle \Big] \nonumber \\
 \times e^{-\frac{i}{\hbar} \int_{t^\prime}^t d\tau [\varepsilon_{n^\prime}({\bf k}^\prime(\tau) + {\bf F} (\tau)\cdot ({\bf R}_{nn}({\bf k}(\tau)) - {\bf R}_{n^\prime n^\prime}({\bf k}^\prime(\tau)))]}.  \;\;\;
\end{eqnarray}

In formally using Eq.~(\ref{OffDiag}) to eliminate the off-diagonal terms in the right-hand side of Eq.~(\ref{MasterEq1}), we obtain a closed form equation for the diagonal elements of $\hat{U}$ alone.
Also, in Eq.~(\ref{OffDiag}) and elsewhere in the analysis, we assume that before the field is turned on
and at $t = t_0$, $\langle n^\prime {\bf K}^\prime|\hat{U}(t_0)|n{\bf K}\rangle = 0$ for $n^\prime {\bf K}^\prime \neq n {\bf K}$ so that the leading term of Eq.~(\ref{OffDiag}) vanishes. Thus, Eq.~(\ref{MasterEq1}) becomes
\begin{subequations}
\begin{eqnarray}\label{MasterEq3}
 i\hbar\frac{\partial}{\partial t} \langle n{\bf K}|\hat{U}|n {\bf K}\rangle = \varepsilon_{n}({\bf k}) \langle n {\bf K}|\hat{U}|n{\bf K}\rangle \nonumber \\
 + \frac{1}{i\hbar} \sum_{n^\prime \neq n} \int_{t_0}^t dt^\prime \Big[\langle n{\bf K}|\hat{U}(t^\prime)|n{\bf K}\rangle - \langle n^\prime{\bf K}|\hat{U}(t^\prime)|n^\prime{\bf K}\rangle \Big] \nonumber \\
 \times\Big\{{\bf F}(t)\cdot{\bf R}_{n^\prime n}({\bf k}(t)) {\bf F}(t^\prime)\cdot {\bf R}_{n n^\prime}({\bf k}(t^\prime))  \nonumber \\
 \times e^{-\frac{i}{\hbar} \int_{t^\prime}^t d\tau [\varepsilon_n({\bf k}(\tau)) + {\bf F}(\tau)\cdot ({\bf R}_{n^\prime n^\prime}({\bf k}(\tau)) - {\bf R}_{nn}({\bf k}(\tau)))]}   \nonumber \\
 + {\bf F}(t)\cdot{\bf R}_{n n^\prime}({\bf k}(t)) {\bf F}(t^\prime)\cdot {\bf R}_{n^\prime n}({\bf k}(t^\prime))
 \nonumber \\
 \times e^{-\frac{i}{\hbar} \int_{t^\prime}^t d\tau [\varepsilon_{n^\prime}({\bf k}(\tau)) - {\bf F}(\tau)\cdot ({\bf R}_{n^\prime n^\prime}({\bf k}(\tau)) - {\bf R}_{nn}({\bf k}(\tau)))]}\Big\},  \;\;\;\;\;\;\;
\end{eqnarray}
or reexpressing Eq.~(\ref{MasterEq3}) as
\begin{equation}\label{MasterEq4}
 i\hbar\frac{\partial}{\partial t} \langle n{\bf K}|\hat{U}|n {\bf K}\rangle = \varepsilon_{n}({\bf k}) \langle n {\bf K}|\hat{U}|n{\bf K}\rangle + J_n(t),
\end{equation}
\end{subequations}
where $J_n(t)$ is the integral term on the right-hand side of Eq.~(\ref{MasterEq3}), we find
\begin{eqnarray}\label{DiagME}
 \langle n{\bf K}|\hat{U}(t)|n {\bf K}\rangle = \frac{1}{i\hbar} \int_{t_0}^t dt^\prime J_n(t^\prime) e^{-\frac{i}{\hbar} \int_{t^\prime}^t d\tau \varepsilon_n({\bf k}(\tau))}  \nonumber  \\
 + \langle n{\bf K}|\hat{U}(t_0)|n {\bf K}\rangle e^{-\frac{i}{\hbar} \int_{t_0}^t d\tau \varepsilon_n({\bf k}(\tau))}. \;\;\;\;
\end{eqnarray}
We note here that for crystal potentials $V_c({\bf r})$ that possess a centrosymmetrical symmetry, ${\bf R}_{n n}({\bf k})$ vanishes identically. Thus, the exponential term in $J_n(t)$ of Eq.~(\ref{MasterEq3}) is greatly simplified so that
\begin{eqnarray}\label{Jn}
 J_n(t) = \frac{1}{i\hbar} \sum_{n^\prime \neq n} \int_{t_0}^t dt^\prime \Big[\langle n{\bf K}|\hat{U}(t^\prime)|n{\bf K}\rangle  \nonumber  \\
 - \langle n^\prime{\bf K}|\hat{U}(t^\prime)|n^\prime{\bf K}\rangle \Big] \nonumber \\
 \times\Big\{{\bf F}(t)\cdot{\bf R}_{n^\prime n}({\bf k}(t)) {\bf F}(t^\prime)\cdot {\bf R}_{n n^\prime}({\bf k}(t^\prime))  \nonumber  \\
 \times e^{-\frac{i}{\hbar} \int_{t^\prime}^t d\tau \varepsilon_n({\bf k}(\tau))}   \nonumber \\
 + {\bf F}(t)\cdot{\bf R}^{\ast}_{n^\prime n}({\bf k}(t)) {\bf F}(t^\prime)\cdot {\bf R}^{\ast}_{n n^\prime}({\bf k}(t^\prime))  \nonumber  \\
 \times e^{-\frac{i}{\hbar} \int_{t^\prime}^t d\tau \varepsilon_{n^\prime}({\bf k}(\tau))}\Big\}.
\end{eqnarray}
We thus conclude that (\ref{DiagME}) and (\ref{Jn}) represent the system of coupled self-consistent linear integral equations for the diagonal elements of $\hat{U}(t)$.

Further, in considering the so-called short-time approximation, we let $\hat{U}_{n{\bf K}n{\bf K}}(t_0) = \delta_{n,n_0} \delta_{{\bf K},{\bf K}_0}$, with $\hat{U}_{n{\bf K}n{\bf K}}(t) \ll 1$ if $n{\bf K} \neq n_0 {\bf K}_0$; and, if we let $\hat{U}_{n^\prime{\bf K}^\prime n^\prime{\bf K}^\prime}(t_0) = \delta_{n^\prime,n_0} \delta_{{\bf K^\prime},{\bf K}_0}$, then the term $J_n(t)$ in Eq.~(\ref{Jn}) becomes
\begin{eqnarray}\label{Jn1}
 J_n(t) = -\frac{1}{i\hbar} \Big\{ {\bf F}(t)\cdot{\bf R}_{n_0 n}({\bf k}_0(t))  \nonumber  \\
 \times \int_{t_0}^t dt^\prime {\bf F}(t^\prime)\cdot {\bf R}^{\ast}_{n_0 n}({\bf k}_0(t^\prime))
 e^{-\frac{i}{\hbar} \int_{t^\prime}^t d\tau \varepsilon_n({\bf k}_0(\tau))}  \nonumber \\
 + {\bf F}(t)\cdot{\bf R}^{\ast}_{n_0 n}({\bf k}_0(t))  \nonumber  \\
 \times \int_{t_0}^t dt^\prime {\bf F}(t^\prime)\cdot {\bf R}_{n_0 n}({\bf k}_0(t^\prime))
 e^{-\frac{i}{\hbar} \int_{t^\prime}^t d\tau \varepsilon_{n_0}({\bf k}_0(\tau))} \Big\}.  \;\;\;\;
\end{eqnarray}
Letting $G_{n_i}(t)$ from Eq.~(\ref{Jn1}) be described by
\begin{subequations}
\begin{equation}\label{G}
  G_{n_i}(t) = \frac{1}{i\hbar} \int_{t_0}^t dt^\prime {\bf F}(t^\prime)\cdot {\bf R}_{n_0 n}({\bf k}_0(t^\prime))
 e^{\frac{i}{\hbar} \int_{t_0}^t d\tau \varepsilon_{n_i}({\bf k}(\tau))},
\end{equation}
where $n_i = n, n_0$, then $J_n(t)$ can be expressed as
\begin{equation}\label{Jn2}
 J_n(t) =  i\hbar\left(\frac{dG_n}{dt}G_n^\ast + \frac{dG_{n_0}^\ast}{dt}G_{n_0} \right).
\end{equation}
\end{subequations}
The differential equation for $\langle n{\bf K}|\hat{U}(t)|n{\bf K}\rangle$ of Eq.~(\ref{MasterEq4}) can be written in a standard form as
\begin{eqnarray}\label{DiffEq}
 i\hbar\frac{\partial}{\partial t} \langle n{\bf K}|\hat{U}(t)|n {\bf K}\rangle - \varepsilon_{n}({\bf k}(t)) \langle n {\bf K}|\hat{U}(t)|n{\bf K}\rangle  \nonumber  \\
 = i\hbar \left(\frac{dG_n}{dt}G_n^\ast + \frac{dG_{n_0}^\ast}{dt}G_{n_0} \right).
\end{eqnarray}
Thus, it follows from Eq.~(\ref{DiffEq}) that for short-time situation, as noted above Eq.~(\ref{Jn1}), we have
\begin{eqnarray}\label{DiagonalEl}
 \langle n{\bf K}|\hat{U}(t)|n {\bf K}\rangle = \frac{1}{i\hbar} \int_{t_0}^t dt^\prime \left(\frac{dG_n}{dt^\prime}G_n^\ast + \frac{dG_{n_0}^\ast}{dt^\prime}G_{n_0} \right)  \nonumber  \\
 \times e^{-\frac{i}{\hbar} \int_{t_0}^{t^\prime} d\tau \varepsilon_n({\bf k}(\tau))},  \;\;\;\;\;
\end{eqnarray}
where $G_{n_i}(t)$ is given in Eq.~(\ref{G}).

Thus, the complete characterization of the diagonal and off-diagonal matrix elements, including the short-time behavior, is determined in terms of the basis of the instantaneous eigenstates. It follows that the time evolution operator in conjunction with the wave functions of Eqs.~(\ref{WF}) and (\ref{Utt0}), with the insertion of complete sets of instantaneous eigenstates, can be used to analyze the time-dependent behavior of the wave function using $\hat{U}(t,t_0)$ in a fashion similar to that done in Sec. III.

\end{document}